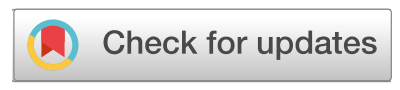

OPEN

# Individual and collective gains from cooperation and reciprocity in a dynamic-network Prisoner's Dilemma driven by extraversion, openness, and agreeableness

David Abián[1,2]✉, Jorge Bernad[1,2,3], Sergio Ilarri[1,2,3] & Raquel Trillo-Lado[1,2]

**How do stable personality differences shape cooperation when social ties can form and dissolve? We model a repeated Prisoner's Dilemma on an endogenous network in which three continuous Big Five traits map to transparent local mechanisms: Extraversion sets a target number of partners, Openness determines how broadly agents search beyond friends-of-friends, and Agreeableness sets a baseline willingness to cooperate. At each encounter, agents combine this baseline with the partner's directly observed history; there are no trait labels, gossip, or global reputations. Ties form when agents are under-connected and are cut when they become over-connected, with cuts prioritising partners who have defected more often. We vary network size (N=30–200), population composition, and the balance between trait-driven and history-driven behaviour. Three robust patterns emerge. First, cooperate first, then reciprocate—high initial willingness to cooperate combined with history-sensitive response—produces systems that are simultaneously more prosperous, fairer, and safer. Second, personality has predictable conditional effects: Agreeableness helps when history matters but hurts when behaviour is mostly trait-driven; Extraversion amplifies the environment; Openness has little net payoff effect. Third, the network reorganises accordingly: degree assortativity stays near zero, whereas agreeable agents increasingly connect to one another when cooperation takes hold.**

Cooperation in decentralised social systems depends on who meets whom, how relationships evolve, and how individuals adjust behaviour based on partners' past actions[1–3]. Repeated exchange can create value, but it also invites exploitation when defection goes unpunished, as formalised in the Prisoner's Dilemma (PD) and its repeated variants[2,3]. Both network structure and the rules governing tie formation and termination shape exposure to partners, opportunities for reciprocity, and the spread of harm[4–7]. Explaining when cooperative systems emerge, how fair they are, and who benefits therefore requires understanding how stable individual differences interact with these dynamics[8,9].

Much computational work fixes the network or treats partner choice as an exogenous heuristic[4–6,10–12]. Related work has shown that cooperation can change when strategic behaviour alters the conditions of future interaction, for example through feedback between individual decisions and environmental risk[13,14]. Our model follows this general idea, but here the changing environment is the social network itself: agents form and cut ties, and these changes affect future opportunities for reciprocity. Many models also assume homogeneous agents or compress person-level differences into a few strategy types[10,15–17]. These simplifications overlook a potentially important source of structure: heterogeneous but persistent dispositions that influence how many relationships people maintain, how broadly they search for partners, and how readily they cooperate before—and after—gaining experience with each partner.

The Prisoner's Dilemma[18] is a two-player, simultaneous-move game in which each player chooses whether to *cooperate* (pursue mutual benefit) or *defect* (act in self-interest). A defector earns $T$ if the partner cooperates and $P$ if the partner defects, whereas a cooperator earns $R$ or $S$, respectively, with $T > R > P > S$ (often also $2R > T + S$)[2,3,17]. Because defection yields a higher payoff regardless of the partner's move, cooperation is individually irrational in the one-shot game even though it maximises joint payoff. In a finitely repeated game with a known last round, backward induction implies defection in every round as the unique subgame-perfect equilibrium[19]. In contrast, in infinite or *indefinite*-horizon settings—where play continues with positive probability—no dominant strategy exists, and the Folk Theorem implies that cooperation can be sustained in equilibrium when the continuation probability is sufficiently high, for example via reciprocal strategies such as tit-for-tat or grim-trigger[20]. Laboratory evidence further shows that people adapt to partners' past actions and differ systematically in baseline cooperativeness and trust-related preferences[21–27].

[1]Departamento de Informática e Ingeniería de Sistemas, Universidad de Zaragoza, 50018 Zaragoza, Spain. [2]Instituto de Investigación en Ingeniería de Aragón (I3A), Universidad de Zaragoza, 50018 Zaragoza, Spain. [3]Jorge Bernad and Sergio Ilarri contributed equally to this work. ✉email: abian@unizar.es

In this work, we focus on a networked Prisoner's Dilemma[11,28]. Agents are nodes connected by undirected edges representing bilateral relationships. In each turn, every connected pair simultaneously plays a one-shot Prisoner's Dilemma. Links are adaptive: agents can unilaterally sever existing ties, which stops future interactions on that edge, and non-neighbouring agents that enter one another's discovery pools may form new ties, enabling future interactions.

To organise individual differences, we use the Big Five (OCEAN) personality framework[8,29,30], a common and empirically robust taxonomy of personality dimensions whose broad structure replicates across instruments, observers, and cultures, and whose traits show substantial rank-order stability across the lifespan[31–35]. We focus on Extraversion, Openness, and Agreeableness because they map most directly onto the social mechanisms studied here, and this mapping is empirically motivated rather than purely ad hoc. Extraversion has been linked to larger and more actively maintained personal networks[36,37]; Agreeableness to prosociality, cooperation, and trust-related behaviour[9,38]; and Openness to curiosity, novelty seeking, network diversity, and greater turnover in social contacts[37,39,40]. Accordingly, in our model Extraversion ($E$) sets an agent's preferred number of concurrent partners (social exposure), Openness ($O$) governs how much partner search extends beyond friends-of-friends to strangers (search breadth), and Agreeableness ($A$) provides a baseline willingness to cooperate (prosocial disposition). Behaviour at each interaction blends this disposition with the partner's own track record with the agent. No global reputations, gossip, public scores, or trait labels are available, so information is strictly local and the model isolates direct reciprocity under local information. We impose this local-information restriction deliberately to isolate the effects of direct reciprocity and partner choice without adding other indirect social-learning channels. Relationships usually form when agents are under-connected and are severed once they become over-connected, with priority given to partners who have defected more often. Memories of how a partner behaved carry over even if a relationship is temporarily broken and later re-formed.

We study this mechanism in pre-specified simulations that vary (i) network size (from 30 to 200), (ii) the population's trait mix (e.g., more or fewer agreeable or extraverted agents), and (iii) the extent to which choices reflect stable disposition versus partner-specific history. We evaluate outcomes at both the individual level (cumulative payoffs) and the collective level (average payoff, inequality, the share of agents harmed, and whether less agreeable agents systematically outperform more agreeable ones). Across conditions, average edge density stabilises quickly after the first few turns and then remains approximately constant, so differences in performance are driven less by changes in overall density than by how ties are reallocated (who interacts with whom) and by the resulting patterns of Prisoner's Dilemma play on those ties.

Prior work has linked personality to behaviour in the Prisoner's Dilemma[41,42], yet how multiple traits jointly shape partner choice and network evolution remains underexplored. Using an agent-based model, we examine how Extraversion, Openness, and Agreeableness govern cooperation and the endogenous formation and cutting of ties. The Results section shows how decision regime and population composition shape individual payoffs, collective performance, and emerging network structure. Formal definitions and implementation details are provided in the Methods.

## Methods

The positive-part operator is $(x)_+ = \max\{x, 0\}$. For any real $x$ and bounds $a \leq b$, we write $[x]_a^b$ for clipping to $[a, b]$. When both agent and time appear, the agent is a subscript and time is an argument, e.g. $k_i(t)$ and $N_i(t)$.

### Turn schedule

At each turn $t \in \{1, \ldots, H\}$, with current graph $G_t = (V, E_t)$:

1. Compute each agent's neighbour set $N_i(t)$ and degree $k_i(t) = |N_i(t)|$.
2. Build discovery pools $\mathcal{P}_i(t)$; attempt *tie formation* for eligible pairs using Eq. 3.
3. Recompute $N_i(t)$ and $k_i(t)$; execute *tie cutting* on existing edges using Eq. 4.
4. On surviving edges, agents play the Prisoner's Dilemma; actions are drawn from the cooperation model in Eq. 2. Update directed dyadic memories $(n^{C}_{\cdot\to\cdot}, n^{D}_{\cdot\to\cdot})$ and cumulative payoffs. Log per-agent metrics, Spearman correlations (payoff vs. metrics), assortativity (degree and numeric attributes), and network statistics.

Newly formed edges may be cut later in the same turn before play, because cutting precedes the interaction step.

### Agents, traits, and scenarios

We simulate $N$ agents as nodes of a simple undirected graph $G_t = (V, E_t)$ over turns $t = 1, \ldots, H$. Each agent $i$ has continuous traits

$$E_i \text{ (Extraversion)}, \quad O_i \text{ (Openness)}, \quad A_i \text{ (Agreeableness)} \quad \in [0, 1].$$

Traits are i.i.d. unless otherwise specified. In the *balanced* condition all three follow $\mathrm{Beta}(5,5)$. In scenario conditions $\{\mathrm{extraversion_hi}, \mathrm{extraversion_lo}, \mathrm{openness_hi}, \mathrm{openness_lo}, \mathrm{agreeableness_hi}, \mathrm{agreeableness_lo}\}$ the named trait uses $\mathrm{Beta}(7.5, 2.5)$ (high) or $\mathrm{Beta}(2.5, 7.5)$ (low); the others remain $\mathrm{Beta}(5,5)$.

Ideal degree (Extraversion). Extraversion maps to agent *i*'s target number of concurrent partners,

$$D_i^\star \;=\; d_{\min} + E_i\,(d_{\max} - d_{\min}), \qquad d_{\min}{=}1,\; d_{\max}{=}10, \tag{1}$$

used by both formation and cutting.

## Network initialisation

We initialise with a bounded-degree graph: each node starts with degree in $\{1,2\}$. We first pair nodes to ensure degree $\geq 1$, then add edges between degree-1 nodes (avoiding multi-edges) until the fraction of degree-2 nodes reaches $f_2 = 0.5$ (with an odd-$N$ safeguard). This yields a sparse, narrow-degree baseline.

## Interaction game and cooperative behaviour

On each surviving edge $\{i,j\}$ at turn *t*, actions are drawn independently given the cooperation probabilities below.

Behavioural policy. Let $n^C_{j\to i}(t)$ and $n^D_{j\to i}(t)$ be the counts of times *j* cooperated with or defected against $i$ up to turn $t{-}1$. The probability that *i* cooperates with *j* at time *t* is a weighted blend of *i*'s Agreeableness and a dyad-specific posterior encoding *i*'s experience interacting with *j*:

$$p\big(C_i \mid j,t\big) \;=\; \beta\, A_i \;+\; (1-\beta)\, \underbrace{\frac{n^C_{j\to i}(t) + sA_i}{n^C_{j\to i}(t) + n^D_{j\to i}(t) + s}}_{\text{dyad-specific posterior}}, \tag{2}$$

with smoothing strength $s = \frac{1}{2}$. The posterior term is the mean of a Beta–Bernoulli model with $sA_i$ pseudo-cooperations and $s(1{-}A_i)$ pseudo-defections. With no history ($n^C{=}n^D{=}0$), the first move equals the disposition ($p(C_i \mid j,t){=}A_i$) for any $\beta$. We study $\beta \in \{0.2, 0.5, 0.8\}$ as representative history-dominant, balanced, and trait-dominant decision regimes, chosen to span the trait–history continuum while keeping both components of Eq. 2 active in the main simulation grid. For completeness, we also simulate the edge cases $\beta = 0.0$ and $\beta = 1.0$; the results are reported in the Supplementary Information. Sensitivity analyses for the global damping parameter $\lambda \in \{0.3, 0.7\}$ are also reported in the Supplementary Information. We treat $\beta$ as a condition-level control parameter rather than as a literal estimate of a fixed human cognitive weight; even when $\beta$ is fixed, the effective influence of dyadic history still increases over time because the posterior is updated after every interaction and the prior is weak.

Payoffs and payoff-asymmetry factor. In a single Prisoner's Dilemma interaction on edge $\{i,j\}$, per-agent outcomes are: *mutual cooperation* (*C*, *C*): both receive *R*; *mutual defection* (*D*, *D*): both receive *P*; *unilateral defection* (*D*, *C*) or (*C*, *D*): the defector receives *T* and the cooperator receives *S*. We use the common normalisation $(T,R,P,S) = (5,3,1,0)$[2,3,17] and subtract a symmetric per-edge cost $c = 2$ each turn. The resulting net one-step payoffs are $T{-}c = +3$, $R{-}c = +1$, $P{-}c = -1$, and $S{-}c = -2$, so being exploited is costly, whereas mutual cooperation is a net gain.

To align partner evaluation with one-step net payoffs, we define the *payoff-asymmetry factor*

$$\eta \;=\; \frac{|S-c|}{R-c} \;=\; 2,$$

and use it to weight defections more heavily than cooperations when prioritising cuts (Eq. 4). Under our payoff normalisation, mutual cooperation yields $R - c = +1$ whereas being exploited yields $S - c = -2$, so one defection is weighted approximately like two cooperations. Thus, $\eta$ is not introduced as a free fitted parameter, but as a quantity induced by the current payoff specification. Under alternative payoff values or edge costs, the corresponding payoff-calibrated $\eta$ could differ. This heavier weighting of defections is also qualitatively consistent with evidence that negative interactions usually carry greater diagnostic weight than positive ones[26,43–45].

Cumulative payoff. Agent *i*'s cumulative payoff through horizon *H* is

$$\Pi_i(H) \;=\; \sum_{t=1}^{H} \sum_{j\in N_i(t)} \big(\pi_{ij}(t) - c\big),$$

where $\pi_{ij}(t)$ is *i*'s PD payoff against *j* at turn *t* before cost. Costs apply only to edges that actually play at *t*.

## Tie dynamics

*Discovery and formation (Openness)*

Let $N_i(t)$ be *i*'s neighbours, $k_i(t) = |N_i(t)|$. Define the friends-of-friends (FoF) set

$$\mathrm{FoF}_i(t) = \Big\{\ell :\; \exists\, j \in N_i(t) \text{ s.t. } \ell \in N_j(t),\; \ell \notin N_i(t),\; \ell \neq i\Big\},$$

and the outsiders set $\mathrm{Out}_i(t) = (V \setminus (N_i(t) \cup \{i\})) \setminus \mathrm{FoF}_i(t)$ with sizes $m{=}|\mathrm{FoF}_i(t)|$ and $M{=}|\mathrm{Out}_i(t)|$. Openness $O_i$ controls the number of outsiders $u_i(t)$ added to the discovery pool:

$$u_i(t) = \begin{cases} 0, & M = 0 \text{ (edge case: no outsiders to choose from)}, \\ \left[\left\lceil \frac{O_i}{1-O_i}\, m \right\rceil\right]_1^{\lceil M/2 \rceil}, & M > 0,\ m > 0,\ O_i < 1 \text{ (general case: contact an outsider share relative to } m\text{)}, \\ \left\lceil \frac{1}{2}\, O_i\, M \right\rceil, & \text{otherwise (edge case:} m = 0 \text{or} O_i = 1, \text{so contact as many outsiders as possible)}, \end{cases}$$

so the outsider proportion increases with $O_i$, at least one outsider is considered when available, and at most half of all outsiders are queried per turn. The discovery pool is

$$\mathcal{P}_i(t) = \mathrm{FoF}_i(t) \;\cup\; \mathrm{sample}(\mathrm{Out}_i(t),\, u_i(t))\,,$$

sampling outsiders uniformly without replacement.

Discovery can be *unilateral*: a non-adjacent pair becomes eligible if at least one endpoint listed the other,

$$\mathcal{C}(t) = \{\{i,j\} \notin E_t :\ j \in \mathcal{P}_i(t) \ \text{ or } \ i \in \mathcal{P}_j(t)\}\,, \qquad L_i(t) = |\{\{i,\ell\} \in \mathcal{C}(t)\}|.$$

By construction, if $\{i,j\} \in \mathcal{C}(t)$, then $L_i(t) \geq 1$ and $L_j(t) \geq 1$.

Each eligible pair attempts to form an undirected tie with probability

$$p_{ij}^{\mathrm{add}}(t) \;=\; p_{ji}^{\mathrm{add}}(t) \;=\; \left[\,\lambda \left(\frac{D_i^\star - k_i(t)}{L_i(t)} + \frac{D_j^\star - k_j(t)}{L_j(t)}\right) \frac{1}{2}\,\right]_0^1, \tag{3}$$

which spreads each endpoint's unmet-degree pressure over the opportunities available, averages the endpoints' propensities (a simple negotiation), and uses a global damping $\lambda \in [0,1]$ to pace formation.

*Cutting and partner evaluation*

Let $n^C_{j\to i}(t)$ and $n^D_{j\to i}(t)$ be the counts, up to turn $t{-}1$, of *j* cooperating with or defecting against *i*. We write $i \leftarrow j$ to emphasise that the evidence concerns *j*'s behaviour toward *i*. Define a per-dyad *defection-risk score* with a Laplace pseudocount:

$$\omega_{i\leftarrow j}(t) = \frac{\eta\, n^D_{j\to i}(t)}{\eta\, n^D_{j\to i}(t) + n^C_{j\to i}(t) + 1}, \qquad \eta = \frac{|S-c|}{R-c} = 2.$$

Thus, under the present payoff normalisation, a single defection is weighted approximately like two cooperations, aligning the cutting rule with the one-step net payoff asymmetry.

Leniency $\delta \in [0,1)$ defines a *soft ideal degree* $(1{-}\delta)D_i^\star$, so cutting pressure activates only above this threshold. Using the same global damping parameter $\lambda$,

$$\mathrm{excess}_i(t) = \lambda\,(k_i(t) - (1{-}\delta)D_i^\star)_+\,.$$

When $k_i(t)$ exceeds the soft ideal, *i* cuts neighbour *j* with probability

$$p_{i\to j}^{\mathrm{cut}}(t) = \left[\,\mathrm{excess}_i(t) \cdot \frac{\omega_{i\leftarrow j}(t)}{\sum_{\ell \in N_i(t)} \omega_{i\leftarrow \ell}(t)}\,\right]_0^1. \tag{4}$$

If $\sum_{\ell \in N_i(t)} \omega_{i\leftarrow\ell}(t) = 0$ (no defection evidence anywhere), no desertion occurs at *i* that turn even if $k_i(t) > (1{-}\delta)D_i^\star$. An undirected edge $\{i,j\}$ is removed if either endpoint's cut draw succeeds. Because desertion draws are independent across neighbours, $k_i$ can stochastically fall below $(1{-}\delta)D_i^\star$ within a turn. Directed dyadic memories persist across cuts and re-formations. There are no global reputations or trait labels. Agents do not receive gossip, public scores, or third-party summaries of others' past behaviour. We adopt this strictly local-information design deliberately to isolate direct reciprocity under local knowledge and to keep partner evaluation tied only to what one agent has directly experienced from another. Any reputational filtering therefore arises only indirectly through local partner choice and the network structure it creates.

## Outcomes and measurements

*Collective performance*

We consider: (i) whether low-*A* agents outperform high-*A* agents, based on the average Spearman correlation between Agreeableness and payoff; (ii) average payoff per agent; (iii) payoff inequality; and (iv) harm prevalence.

*Payoff inequality* $I(t) \in [0,1]$ is the normalised-by-absolute-mean Gini of cumulative payoffs at turn $t$,

$$I(t) \;=\; \begin{cases} 0, & N \leq 1 \text{ or } \sum_{i=1}^N |\Pi_i(t)| = 0, \\ \frac{\sum_{i=1}^N \sum_{j=1}^N |\Pi_i(t) - \Pi_j(t)|}{2(N-1)\sum_{i=1}^N |\Pi_i(t)|}, & \text{otherwise.} \end{cases}$$

This index is scale-invariant, accommodates negative payoffs, equals 0 under perfect equality, and approaches 1 under maximal dispersion.

*Harm prevalence* $h(t) \in [0, 1]$ is the share of agents with negative cumulative payoff at turn $t$,

$$h(t) \;=\; \frac{1}{N}\sum_{i=1}^{N} 1\{\Pi_i(t) < 0\}\,.$$

*Correlations and assortativity*

At each turn we compute Spearman's $\rho$ between cumulative payoff and each trait across agents. Degree assortativity is computed with Newman's degree assortativity coefficient, and attribute assortativity for Agreeableness with the numeric assortativity coefficient[46,47].

## Experimental design

We run a full factorial grid:

$$N \in \{30,\, 100,\, 200\}, \quad H = 300, \quad \beta \in \{0.2,\, 0.5,\, 0.8\}, \quad \text{seeds 101–120} \;\; \text{(20 replicates per cell)}.$$

Robustness runs with $\beta \in \{0.0, 1.0\}$ are reported in the Supplementary Information. Networks use the bounded-degree initialisation ($f_2$=0.5). Baseline traits are $\mathrm{Beta}(5, 5)$ with scenario variants as above. Payoffs are $(T, R, P, S) = (5, 3, 1, 0)$ with per-edge cost $c = 2$. Formation and cutting both use $\lambda = 0.5$; sensitivity runs varying the global damping parameter to $\lambda \in \{0.3, 0.7\}$ use the same settings otherwise and are also reported in the Supplementary Information. Cutting leniency is $\delta = 0.2$. Table 1 summarises all settings.

## Aggregation and uncertainty

For each condition (fixed $N$, $\beta$, scenario), we average seed-level statistics per turn and report 95% CIs using a $t$-based $\mathrm{SD}/\sqrt{n}$ estimator across seeds:

$$\bar{z} = \frac{1}{n}\sum_{s=1}^{n} z_s, \qquad \mathrm{CI}_{95} = t_{0.975,\, n-1} \cdot \frac{\mathrm{SD}(z_1, \ldots, z_n)}{\sqrt{n}}.$$

Thus, all reported effects are ensemble-level tendencies: they summarise averages across seeds (and, where applicable, sizes), whereas individual runs of the same condition can behave differently due to stochasticity in traits, meetings, and game outcomes. For Fig. 1, time points are first averaged over sizes $N \in \{30, 100, 200\}$ within each seed, then CIs are computed across seeds only; side-bar end-state metrics are averaged across seeds and then across sizes.

## Reproducibility and implementation

Each run initialises a NumPy RNG with the provided seed; per-run parameters and library versions (NumPy, pandas, NetworkX, SciPy) are saved with outputs. All results are written as CSV files. The raw results, source code, and command-line invocation are cited in the Code and data availability statement.

## Use of generative AI tools

The authors used a large language model (ChatGPT, OpenAI; accessed November 2025) to assist with language editing and with drafting and refactoring source code. All model outputs (text and code) were reviewed, verified, and edited by the authors, who remain fully responsible for the content of the article. No generative AI tools were used for the conception or design of the study, for data generation, or for the interpretation of results, and no human participant data were processed by these tools. All final code is archived in the public repository cited in the Code and data availability statement.

| | |
|---|---|
| Agents $N$ | $30,\ 100,\ 200$ |
| Turns $H$ | 300 |
| Seeds per condition | 20 (IDs 101–120) |
| Trait–history mixing weight $\beta$ | $0.2,\ 0.5,\ 0.8$ |
| Trait distributions | Baseline: Beta(5, 5) for $E$, $O$, $A$;<br>Scenario skew: Beta(7.5, 2.5) (high) or Beta(2.5, 7.5) (low) for one trait |
| Ideal degree | Eq. 1, $d_{\min} = 1, d_{\max} = 10$ |
| PD payoffs | $(T, R, P, S) = (5,\ 3,\ 1,\ 0)$ |
| Per-edge cost | $c = 2$ (symmetric) |
| Global damping | $\lambda = 0.5$ (applies to formation and cutting) |
| Leniency | $\delta = 0.2$ (soft slack below ideal degree) |
| Initialisation | Bounded-degree: degree $\in \{1, 2\}$, fraction degree-2 $f_2 = 0.5$ |

**Table 1**. Summary of parameter values and simulation settings for the main analyses. Robustness analyses with $\beta \in \{0, 1\}$ and $\lambda \in \{0.3, 0.7\}$ are reported in the Supplementary Information.

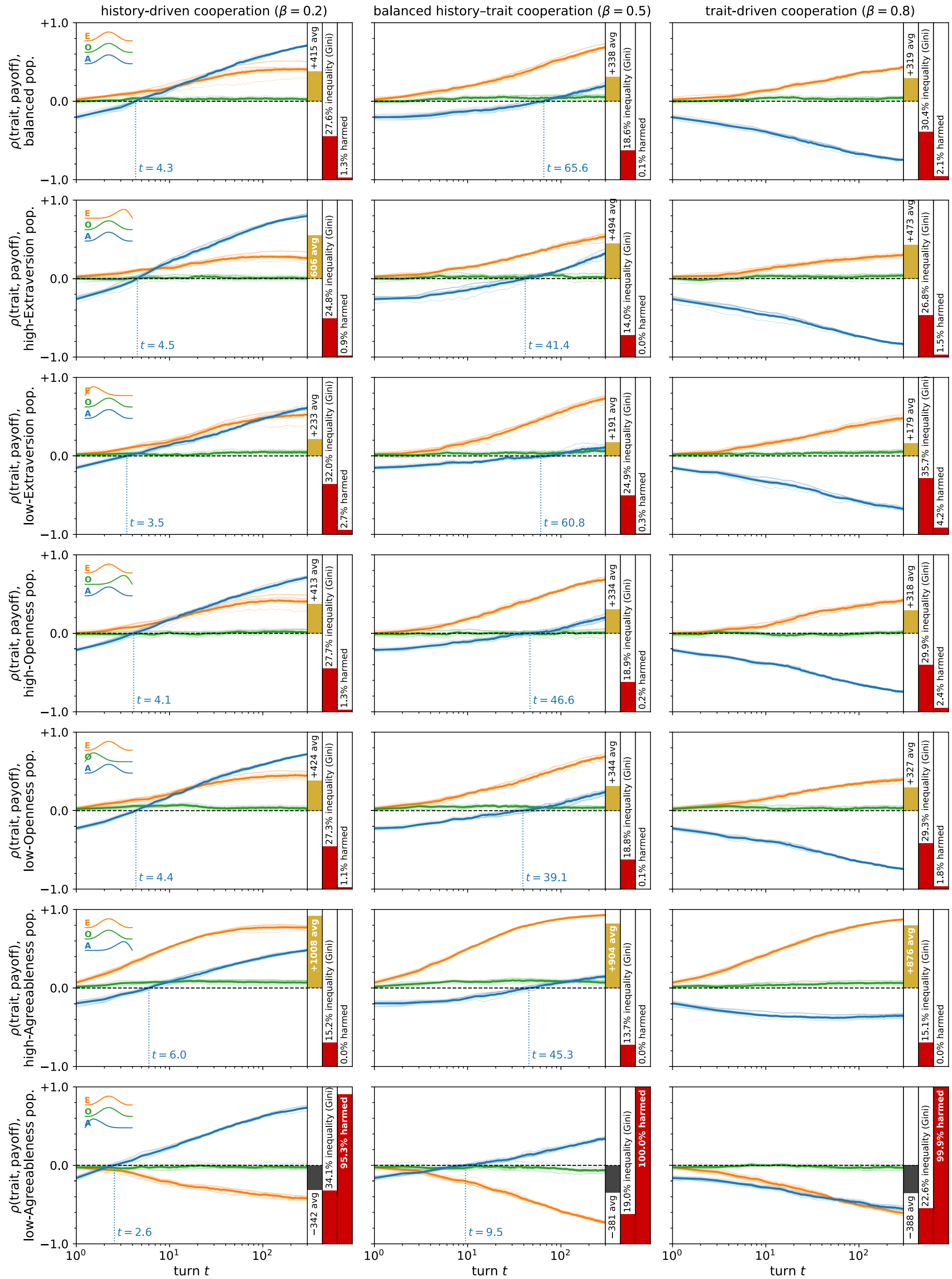

**Fig. 1**. Trait–payoff correlations over time across scenarios and decision rules. Rows vary the population trait scenario (baseline; high/low Extraversion, Openness, Agreeableness). Columns vary the mixing weight $\beta$ that blends Agreeableness with dyad-specific history in cooperation decisions (history-dominant, balanced, trait-dominant). Within each panel, thick lines plot mean Spearman correlation $\rho(\mathrm{trait}, \mathrm{payoff})$ across seeds and sizes; shaded ribbons show $95\%$ confidence intervals (CI). Thin lines show size-conditional means by $N$ to illustrate size robustness. The horizontal dashed line marks $\rho = 0$; when the Agreeableness–payoff curve first crosses 0, a vertical dotted line marks the turn. Side bars summarise end-state ($t = H = 300$) properties averaged across seeds and sizes: average payoff per agent (gold if positive, grey if negative), payoff inequality (normalised-by-absolute-mean Gini), and harm prevalence (share of agents with negative cumulative payoff). Insets depict the trait distributions used in each row (balanced: $\mathrm{Beta}(5, 5)$; scenario rows skew one trait to $\mathrm{Beta}(7.5, 2.5)$ high or $\mathrm{Beta}(2.5, 7.5)$ low). See Methods for parameter grid, payoff normalisation, and definitions of metrics.

## Results

Across $N \in \{30, 100, 200\}$, cooperation trajectories and end-state outcomes are nearly size-robust; the principal moderators are $\beta$ (cooperation based on trait or history) and the population's $A_i$ distribution, not network scale.

### Extraversion (*E*)

Extraversion is generally beneficial from the outset for both individuals and collectives across scenarios. However, in particularly low-$A$ populations, which produce hostile systems in which interaction is more harmful than isolation, Extraversion becomes maladaptive, since social engagement amplifies exposure to exploitation (e.g. $\rho = -0.737 \pm 0.037$ at $t = 300$ for $\beta = 0.5$ and low-$A$ $N = 100$, with $-381.9$ payoff/agent at $t = 300$, harm prevalence $h(300) = 100.0\%$).

Extraversion is especially profitable at an individual level in more cooperative populations, where the expected payoff of each extra partner is higher (e.g. $\rho = 0.931 \pm 0.008$ at $t = 300$ for $\beta = 0.5$ and high-$A$ $N = 100$); in less extraverted populations (e.g. $\rho = 0.772 \pm 0.024$ at $t = 300$ for $\beta = 0.5$ and low-$E$ $N = 100$, versus $\rho = 0.659 \pm 0.037$ for balanced and $\rho = 0.465 \pm 0.030$ for high-$E$); and when cooperative or defective actions are determined by a balance between partner-specific adaptation based on dyadic history and Agreeableness (or lack thereof) as a stable trait (e.g. $\rho = 0.659 \pm 0.037$ at $t = 300$ for $\beta = 0.5$ and balanced $N = 100$, versus $\rho = 0.415 \pm 0.033$ for $\beta = 0.2$ and $\rho = 0.424 \pm 0.038$ for $\beta = 0.8$).

High-$E$ populations also tend to be both more collectively profitable and less unequal (e.g. $+460.4$ payoff/agent at $t = 300$, harm prevalence $h(300) = 0.0\%$, payoff inequality $I(300) = 0.147$ for $\beta = 0.5$ and high-$E$ $N = 100$) than balanced ($+330.4$ payoff/agent, harm prevalence $h(300) = 0.0\%$, payoff inequality $I(300) = 0.183$) and low-$E$ populations ($+203.0$ payoff/agent, harm prevalence $h(300) = 0.1\%$, payoff inequality $I(300) = 0.233$).

Group size also appears to moderate the benefits of Extraversion. When Extraversion is beneficial and cooperation is sufficiently history-driven, Extraversion is more individually beneficial in larger populations (e.g. $\rho = 0.506 \pm 0.031$ at $t = 300$ for $\beta = 0.2$ and balanced $N = 200$, versus $\rho = 0.415 \pm 0.033$ for $N = 100$, and $\rho = 0.291 \pm 0.081$ for $N = 30$). A plausible interpretation is that larger populations make additional social exposure more valuable by enlarging the pool of feasible replacement partners and reducing dependence on any single tie. Under history-sensitive reciprocity, this may help highly extraverted agents convert greater exposure into more net-positive relationships.

Degree assortativity tends to be near zero or slightly negative, especially in the smallest systems (see Table 2).

### Openness (*O*)

Openness is largely neutral, with only modest effects on individual payoffs. Across conditions, correlations range narrowly from $\rho = -0.101 \pm 0.047$ (low-$A$ $N = 100$, $\beta = 0.5$) to $\rho = 0.099 \pm 0.030$ (low-$E$ $N = 200$, $\beta = 0.5$). Small benefits appear in high-$A$ populations (e.g. $\rho = 0.095 \pm 0.050$ at $t = 300$ for $\beta = 0.5$ and high-$A$ $N = 100$), where outsider encounters are relatively cooperative and modestly broaden useful contacts. By contrast, in low-$A$ populations, higher Openness is slightly detrimental (e.g. $\rho = -0.101 \pm 0.047$ for $\beta = 0.5$ and low-$A$ $N = 100$), since maintaining trusted long-term ties and reliable partner information is crucial, and frequent outsider contact undermines stability.

### Agreeableness (*A*)

The individual benefits of Agreeableness, when they arise, are typically delayed (see Table 3). In high-$A$ populations it takes longer for unusually agreeable agents to be identified and rewarded, so the onset of a positive $A$–payoff association occurs later (e.g., 45.3 turns on average for $\beta = 0.5$ and high-$A$ populations) than in low-$A$ populations (e.g., 9.5 turns on average for $\beta = 0.5$ and low-$A$ populations). By contrast, when behaviour is predominantly trait-driven (large $\beta$) and dyad-specific learning is weak, relative Agreeableness remains individually costly at every horizon we observe (e.g., $\rho = -0.748 \pm 0.024$ at $t = 300$ for $\beta = 0.8$, balanced $N = 100$), including in populations with high overall $A$ (e.g., $\rho = -0.332 \pm 0.049$ at $t = 300$ for $\beta = 0.8$, high-$A$ $N = 100$) and low overall $A$ (e.g., $\rho = -0.517 \pm 0.057$ at $t = 300$ for $\beta = 0.8$, low-$A$ $N = 100$). Robustness runs sharpen the same contrast (Supplementary Information): at $t = 300$, $\rho(A, \Pi) \in [0.435,\ 0.810]$ for $\beta = 0.0$ but $\rho(A, \Pi) \in [-0.941, -0.620]$ for $\beta = 1.0$ across all scenarios and sizes. The same sign pattern is robust to moderate changes in rewiring pace (Supplementary Information): at $t = 300$, $\rho(A, \Pi)$ remains positive in all 21 size$\times$scenario conditions for $\beta = 0.2$ at both $\lambda = 0.3$ and $\lambda = 0.7$, negative in all 21 conditions for $\beta = 0.8$, and positive in 20 out of 21 conditions for $\beta = 0.5$ at $\lambda = 0.7$.

| N | $\beta$ | Population scenario | average $r \pm$ CI |
|---|---|---|---|
| ... | ... | ... | ... |
| 30 | 0.5 | High-Openness | $-0.050 \pm 0.059$ |
| 30 | 0.2 | Balanced | $-0.052 \pm 0.051$ |
| 30 | 0.5 | Balanced | $-0.053 \pm 0.085$ |
| 30 | 0.2 | Low-Openness | $-0.066 \pm 0.045$ |
| 30 | 0.5 | High-Agreeableness | $-0.067 \pm 0.046$ |
| 30 | 0.8 | High-Agreeableness | $-0.078 \pm 0.052$ |
| 30 | 0.8 | Low-Agreeableness | $-0.091 \pm 0.067$ |

**Table 2.** Systems with average degree assortativity $|r| \geq 0.05$ at $t = 300$.

| On average, $\rho(A, \text{payoff}) > 0$ when... | | | |
|---|---|---|---|
| N | $\beta$ | Population scenario | $t > \ldots$ |
| 100 | 0.2 | Low-Agreeableness | 2.2 |
| 200 | 0.2 | Low-Agreeableness | 2.6 |
| 200 | 0.2 | Low-Extraversion | 3.1 |
| 30 | 0.2 | Low-Extraversion | 3.4 |
| 30 | 0.2 | Low-Agreeableness | 3.5 |
| 200 | 0.2 | High-Extraversion | 3.7 |
| 200 | 0.2 | High-Openness | 3.7 |
| 100 | 0.2 | High-Openness | 3.9 |
| 100 | 0.2 | Low-Extraversion | 3.9 |
| 200 | 0.2 | Balanced | 4.2 |
| ... | ... | ... | ... |
| 100 | 0.5 | High-Agreeableness | 36.6 |
| 100 | 0.5 | Balanced | 43.1 |
| 30 | 0.5 | High-Extraversion | 51.9 |
| 200 | 0.5 | Low-Openness | 55.7 |
| 30 | 0.5 | High-Agreeableness | 58.5 |
| 30 | 0.5 | High-Openness | 64.5 |
| 30 | 0.5 | Balanced | 66.2 |
| 200 | 0.5 | Balanced | 72.4 |
| 100 | 0.5 | Low-Extraversion | 90.5 |
| 200 | 0.5 | Low-Extraversion | 98.9 |
| Any | 0.8 | Any | $+\infty$ |

**Table 3**. Systems in which, on average, it took shorter (top) or longer (bottom) for relative Agreeableness to become an individually beneficial trait, based on results until time horizon $H = 300$.

When cooperation and defection are driven mainly by dyadic history (small $\beta$) or by a trait–history balance (moderate $\beta$), Agreeableness becomes individually advantageous in the shorter or longer run, respectively (e.g., $\rho = 0.718 \pm 0.026$ at $t = 300$ for $\beta = 0.2$, balanced $N = 100$; $\rho = 0.240 \pm 0.051$ at $t = 300$ for $\beta = 0.5$, balanced $N = 100$; see also Table 3). At the collective level, high-$A$ populations achieve the largest average payoffs and the lowest inequality across scenarios (e.g., $+903.0$ payoff/agent at $t = 300$, harm prevalence $h(300) = 0.0\%$, payoff inequality $I(300) = 0.135$ for $\beta = 0.5$, high-$A$ $N = 100$), with the strongest aggregate gains when dyadic history plays the leading role (e.g., $+1011.6$ payoff/agent at $t = 300$, $h(300) = 0.0\%$, $I(300) = 0.153$ for $\beta = 0.2$, high-$A$ $N = 100$).

Agreeableness also becomes assortative over time in high-$A$ populations (see Table 4): agreeable agents preferentially connect to (and benefit) one another while avoiding lower-$A$ partners. The largest Agreeableness assortativity coefficients occur for high-$A$ $N = 100$, $\beta = 0.2$ ($r = 0.174 \pm 0.036$); high-$A$ $N = 200$, $\beta = 0.2$ ($r = 0.174 \pm 0.026$); and high-$A$ $N = 200$, $\beta = 0.5$ ($r = 0.115 \pm 0.030$).

## Discussion

We analysed a dynamic-network Prisoner's Dilemma in which three stable traits affect only local, observable choices: how many partners to keep (exposure; Eq. 1), how broadly to search for partners (via Openness), and how readily to cooperate before learning about a partner (baseline Agreeableness blended with dyadic history; Eq. 2). Ties form by spreading unmet-degree pressure across discovered opportunities (Eq. 3) and are cut once degree exceeds a soft target, prioritising partners with worse defection records (Eq. 4). Per-edge costs create a 2 : 1 loss–gain asymmetry between being exploited and mutually cooperating, and we match this asymmetry when weighting defections in the cutting rule. Average edge density stabilises quickly after the first few turns and then remains approximately constant, so cross-condition differences in performance arise mainly from which ties are retained or replaced (who interacts with whom) and from the resulting patterns of Prisoner's Dilemma play on those ties, rather than from global changes in sparsity.

Two direct mechanistic consequences follow from this design. First, holding degree pressure fixed, an additional defection by *j* strictly increases the probability that *i* cuts edge $\{i, j\}$, whereas an additional cooperation weakly decreases it, via the defection-risk score $\omega_{i \leftarrow j}$ in Eq. 4. Second, because $c > 0$, edges that fail to achieve mutual cooperation on average reduce cumulative payoff; a history-sensitive retention rule therefore reallocates scarce degree towards reciprocators relative to any rule that ignores partner histories. Directed memories persist across cuts and later re-formations, so agents can selectively rebuild relationships with partners who treated them well in the past.

| N | $\beta$ | Population scenario | average $r \pm$ CI |
|---|---|---|---|
| 100 | 0.2 | High-Agreeableness | $0.174 \pm 0.036$ |
| 200 | 0.2 | High-Agreeableness | $0.174 \pm 0.026$ |
| 200 | 0.5 | High-Agreeableness | $0.115 \pm 0.030$ |
| 30 | 0.2 | High-Agreeableness | $0.081 \pm 0.076$ |
| 100 | 0.5 | High-Agreeableness | $0.080 \pm 0.031$ |
| 200 | 0.8 | High-Agreeableness | $0.078 \pm 0.022$ |
| 100 | 0.8 | High-Agreeableness | $0.066 \pm 0.036$ |
| 30 | 0.5 | High-Agreeableness | $0.059 \pm 0.055$ |
| 200 | 0.2 | High-Openness | $0.058 \pm 0.024$ |
| 100 | 0.2 | Balanced | $0.055 \pm 0.043$ |
| ... | ... | ... | ... |
| 30 | 0.2 | Low-Agreeableness | $-0.052 \pm 0.049$ |
| 30 | 0.5 | Low-Extraversion | $-0.094 \pm 0.047$ |

**Table 4.** Systems with average Agreeableness assortativity $|r| \geq 0.05$ at $t = H = 300$.

Across the full grid ($N \in \{30, 100, 200\}$; $\beta \in \{0.2, 0.5, 0.8\}$; trait scenarios), the outcomes in Fig. 1 align with these incentives. Giving more weight to dyadic history (smaller $\beta$) yields higher average payoff, lower harm prevalence, and lower payoff inequality. Trait–payoff associations follow the same logic: Agreeableness benefits individuals when histories are taken seriously (small–moderate $\beta$) but is costly when behaviour is largely trait-driven (large $\beta$); Extraversion amplifies the prevailing environment by increasing exposure to available opportunities, being beneficial in cooperative regimes and harmful in hostile, low-$A$ populations; and Openness has modest, context-dependent effects because it changes whom agents meet, not how they manage existing ties. The same mechanism may also help explain why the payoff advantage of Extraversion is somewhat stronger in larger populations when reciprocity is operative: larger groups offer a broader set of feasible replacement partners, making additional exposure less dependent on any single tie. Consistently, across all conditions, mean payoff per agent is higher at $\beta = 0.0$ than at $\beta = 1.0$ (minimum difference $+144.9$, maximum $+806.0$) and, averaged over net-beneficial regimes (mean payoff $> 0$), $\beta = 0.0$ yields lower or equal harm (1.47% vs. 11.77%) and lower inequality ($I = 0.270$ vs. 0.427) (Supplementary Information). The regularities we emphasise are ensemble-level: they summarise the distribution of outcomes across many runs under a given condition, while individual simulated worlds can deviate—sometimes substantially—due to stochasticity in traits, meetings, and game outcomes. These results connect to a broader literature on feedback-evolving games in which strategy and environment co-determine one another [13,14]. In our case, the endogenous environment is neither a public resource nor a risk level, but the evolving network of bilateral relationships, updated through local partner choice and memory.

Varying the global damping parameter over $\lambda \in \{0.3, 0.7\}$ leaves these sign patterns intact. The main quantitative effect is that larger $\lambda$ produces a sharper early density transient, a lower steady-state degree, and a later emergence of positive $A$–payoff associations when $\beta = 0.5$, without changing the overall advantage of history-sensitive over trait-dominant cooperation (Supplementary Information).

Network structure reflects the same mechanism. Degree assortativity is near zero or slightly negative (Table 2) because formation depends on mutual degree pressure rather than similarity in degree. By contrast, assortativity in Agreeableness emerges in cooperative regimes (Table 4), as agents who more often elicit cooperation are preferentially retained by one another under the history-sensitive cutting rule. In these regimes, more agreeable agents not only fare better on average but also increasingly interact with and benefit from each other, which helps reduce inequality and harm prevalence.

Our claims are intentionally scoped. They are conditional on a deliberately local-information design: agents rely on local dyadic memory, face symmetric per-edge costs $c > 0$, cut partners with probabilities that increase with defection history, and do not use global reputation, third-party punishment, or centralised enforcement. Within this design, these ingredients are sufficient for the joint improvements in prosperity, fairness, and safety that we report; we do not claim that they are necessary beyond it. The model also assumes direct dyadic information only, with no gossip or public reputation, persistent memory with no decay or recency weighting, independent traits drawn from Beta distributions, perfect observation of actions, and fixed decision rules. In particular, $\beta$ does not adapt with relationship age or context, even though the practical influence of history grows as evidence accumulates. Allowing memory to fade or the trait–history balance to adapt could delay, weaken, or in some regimes alter the emergence of positive returns to Agreeableness. Our claims are also conditional on the present payoff specification, from which the defection weight is derived as $\eta = |S - c|/(R - c) = 2$; we did not study alternative payoff matrices or edge costs in the present revision, so we do not claim robustness beyond this calibration.

Several extensions follow naturally. First, additional traits could be incorporated: we focused on Extraversion, Openness, and Agreeableness because they map directly to exposure, search breadth, and baseline cooperativeness under local information; specifying mechanisms for Conscientiousness and Neuroticism would complete the Big Five. Second, preregistered sensitivity analyses could vary payoff parameters ($T$, $R$, $P$, $S$), the per-edge cost $c$, smoothing strength $s$, damping $\lambda$, leniency $\delta$, the outsider cap, and weak inter-trait correlations to test robustness, including whether the same qualitative patterns persist under other payoff specifications and their corresponding payoff-calibrated values of $\eta$. Third, the environment could be enriched by adding reputational spillovers, noisy action observation, memory decay, heterogeneous costs and leniencies, multiplex ties, dynamic payoffs, partially adaptive Extraversion or Openness, alternative formation and cutting rules based on the number of common neighbours, or adaptive trait–history mixing in which $\beta$ changes with relationship duration, accumulated evidence, or situational uncertainty. Fourth, beyond personality, agents could differ in skills/capacities (e.g., learning and memory, inference noise, or strategic sophistication) to assess how aptitude heterogeneity reshapes reciprocity and network evolution. Together, these extensions would help delineate where the "cooperate first, then reciprocate" pattern persists and where personality-linked advantages and disadvantages are reshaped by broader institutional and informational conditions.

## Data availability

All simulation code used to run the experiments and produce the analyses in this article is available on GitHub at https://github.com/davidabian/bigthree-dynet-pd.

## Code availability

The full set of simulation outputs analysed in the paper is archived on Zenodo at https://doi.org/10.5281/zenodo.17714612.

## Author contributions
D.A. conceived the study, conducted the analyses, interpreted the results, designed the figure, and wrote the manuscript. J.B. revised the Methods section, replicated the study based on that section, and confirmed the main findings. S.I. contributed to the writing and added references. R.T.L. led the administrative procedures and fundraising efforts. All authors reviewed the manuscript.

## Funding
This publication is part of the project PID2020-113037RB-I00, funded by MICIU/AEI/10.13039/501100011033, as well as I-2018/026/PIP Modalidad-A UZ. In addition to the NEAT-AMBIENCE project, D.A., S.I., and R.T.L. acknowledge the support of the Departamento de Ciencia, Universidad y Sociedad del Conocimiento del Gobierno de Aragón (Government of Aragon: Group Reference T64_23R, COSMOS research group). J.B. was supported by the I+D+i project PID2020-113903RB-I00 (funded by MCIN/AEI/10.13039/501100011033) and the project T42_23R (funded by Gobierno de Aragón).

## Declarations

## Competing interests
The authors declare no competing interests.

## Ethical approval
This research did not involve human participants, human data, human tissue, or live animals.

## Additional information
**Supplementary Information** The online version contains supplementary material available at https://doi.org/10.1038/s41598-026-49942-w.

**Correspondence** and requests for materials should be addressed to D.A.